\documentclass[aps,prl,floatfix,twocolumn,nofootinbib]{revtex4}
\pdfoutput=1

\usepackage{amssymb}
\usepackage{amsmath}
\usepackage{amsfonts}
\usepackage{graphicx}
\usepackage{color}
\usepackage{xspace}
\usepackage{ulem}
\usepackage{simplewick}

\def\lan   {\langle}
\def\ran   {\rangle}
\def\PhiN {\Phi}
\def\Phichi {\Phi_{\mbox{\tiny{$\chi$}}}}
\def\mPhichi {m_{\mbox{\tiny{$\Phichi$}}}}
\def\mPhiN {m_{\mbox{\tiny{$\PhiN$}}}}
\def\mN {m_{\mbox{\tiny{$N$}}}}

\def\dims{ \chi \chi \nu\nu}

\newcommand{\AddrUCL}{
Department of Physics and Astronomy, University College London,\\
London WC1E 6BT, United Kingdom
}

\begin{document}

\title{Dark matter origins of neutrino masses}

\author{Wei-Chih Huang} \email{wei-chih.huang@ucl.ac.uk}\affiliation{\AddrUCL}
\author{Frank F. Deppisch} \email{f.deppisch@ucl.ac.uk}\affiliation{\AddrUCL}

\begin{abstract}
We propose a simple scenario that directly connects the dark matter (DM) and neutrino mass scales. Based on an interaction between the DM particle $\chi$ and the neutrino $\nu$ of the form $\dims/\Lambda^2$, the DM annihilation cross section into the neutrino is determined and a neutrino mass is radiatively induced. Using the observed neutrino mass scale and the DM relic density, the DM mass and the effective scale $\Lambda$ are found to be of the order MeV and GeV, respectively. We construct an ultraviolet-complete toy model based on the inverse seesaw mechanism which realizes this potential connection between DM and neutrino physics. 
\end{abstract}

\maketitle

\section{Introduction}\label{sec:intro}
The Standard Model~(SM) is not able to explain the existence of Dark Matter~(DM) in the universe as well as the finite masses of neutrinos. Experimentally, both phenomena are firmly established. The two neutrino mass-squared differences are very well measured in neutrino oscillation experiments~\cite{Agashe:2014kda}. Together with the upper limit on the sum of the neutrino masses, $\sum m_{\nu} \lesssim 0.66$ eV, derived from cosmological observations~\cite{Ade:2013zuv}, they imply that the heaviest active neutrino has a mass of 0.05 to 0.22 eV. While the DM mass is largely unconstrained, the crucially important DM relic abundance is very well measured at $\Omega h^2 = 0.12$~\cite{Ade:2013zuv}.

Connections between DM physics and the origin and size of the neutrino masses have been proposed in the literature in the context of radiative neutrino mass models, for example in Refs.~\cite{Krauss:2002px, Ma:2006km, Deppisch:2014jka}, where the neutrino mass is induced radiatively with DM particles and heavy neutrinos in the loop. In these models, the neutrino mass scale depends on the DM and heavy neutrino masses as well as various coupling constants. This implies that the DM mass can not be uniquely determined given the observations, unless other model parameters are fixed. An alternative scenario was proposed in Refs.~\cite{Boehm:2006mi, Farzan:2009ji}. Similar to our case, it connects neutrino physics with an MeV scale DM particle, although the underlying model is quite different.

In this work, we propose a simple scenario that connects the DM particle and neutrino mass scales. Starting with an effective 6-dimensional operator
\begin{align}
\label{eq:theoperator}
	\frac{\dims}{\Lambda^2},
\end{align}
where $\chi$ refers to a gauge singlet Majorana DM particle, while $\nu$ is the SM neutrino\footnote{We neglect the flavour structure of the three neutrinos and work with one Majorana neutrino field with mass scale $m_\nu \approx 0.1$~eV.}. Here and in the following, we use the two-component Weyl spinor notation for all fermionic fields. We implicitly assume that $\chi$ is odd under a $Z_2$ symmetry to ensure its stability. Assuming that this operator is the only one coupling DM to SM particles, the DM annihilation cross section times the DM relative velocity $v_\text{rel}$ is approximated by $ \sigma v_\text{rel} \approx  m^2_\chi / (\pi \Lambda^4)$. This implies a DM relic abundance of $\Omega h^2 \approx 8.2\times 10^{-10}\text{GeV}^{-2} / (\sigma v_\text{rel})$. On the other hand, the neutrino receives a radiative mass by contracting two $\chi$ fields in the interaction operator, $m_{\nu} \approx 55 m^3_\chi/ (\pi^2 \Lambda^2)$. Using the experimental data on the DM relic abundance and the light neutrino mass scale, the DM mass $m_\chi$ and the scale $\Lambda$ of the interaction operator can be determined easily,
\begin{align}
	m_\chi  &\approx 0.4 \text{ MeV}
		\left(\frac{m_\nu}{0.1\text{ eV}}\right)^{1/2} 
		\left(\frac{\Omega h^2}{0.12}\right)^{1/4},\\
	\Lambda &\approx 1.5\text{ GeV} 
		\left(\frac{m_\nu}{0.1 \text{ eV}}\right)^{1/4} 
		\left(\frac{\Omega h^2}{0.12}\right)^{3/8}.
\end{align}
Naturally, $m_\chi$ and $\Lambda$ are of the order MeV and GeV, respectively. The effective operator scale $\Lambda$ is far below the electroweak~(EW) scale, which is why the operator is not invariant under the SM gauge group. It also naturally implies the existence of at least one more particle lighter than the EW scale in order to obtain the interaction operator $\dims / \Lambda^2$. We proceed by constructing a possible model that realizes the previous operator in two steps: firstly by discussing an effective Lagrangian, and then a possible fully ultraviolet (UV)-complete toy model.

\section{Effective Lagrangian}\label{sec:model}
The natural scale of the operator \eqref{eq:theoperator} is GeV and in order to discuss a possible SM effective model we have to introduce another light particle that connects the DM sector with the SM. In addition, we assume that the only source of lepton number violation (LNV), that generates the DM Majorana mass, is situated in the hidden sector. We do not specify this source of LNV but it could for example result from a seesaw-like mechanism in the hidden sector. Note that one has to make sure that in the UV-complete theory, the hidden sector does not couple to the SM directly, i.e. it has to go through the DM particle $\chi$. Therefore, any other effective operators have to {\it conserve} lepton number which for example forbids the Weinberg operator $LHLH$. 

We introduce a complex scalar $\PhiN$ with two units of lepton number, $L(\PhiN) = 2$ which connects the DM and SM sectors,
\begin{align}
	\mathcal{L} \supset c_2 \, \PhiN\chi\chi + \frac{\PhiN^* LH LH}{\Lambda^2_*} + h.c..
\label{eq:PhiN_SM}
\end{align}
Here, $L$ and $H$ are the SM lepton and Higgs boson doublets, respectively. Choosing $L(\chi)=-1$, the Lagrangian \eqref{eq:PhiN_SM} conserves lepton number. After integrating out $\PhiN$ and EW breaking, $H = (0,v)^T$, one obtains
\begin{align}
	\mathcal{L} \supset \frac{\dims}{\Lambda^2} + h.c.,
\label{eq:effective_L}
\end{align}
where $\Lambda =  \Lambda_* \mPhiN / ( \sqrt{c_2} v)$.

\begin{figure}
\centering
\includegraphics[width=\columnwidth]{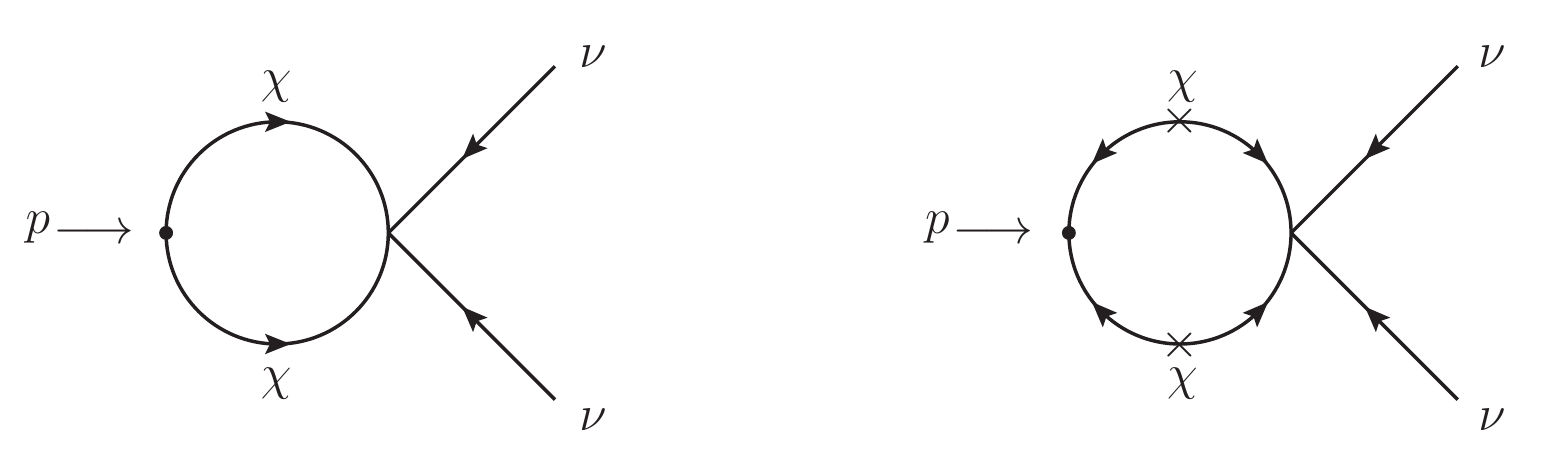}
\caption{Loop diagrams generating a Majorana neutrino mass. The arrow represents chirality; if the arrow direction is the same as that of the momentum, it represents the left-handed chirality. One of the contributions~(right panel) involves a chirality flip.}
\label{fig:mass_gen}
\end{figure} 
Due to the Majorana nature of the DM particle $\chi$, the light neutrino $\nu$ will obtain a loop-induced Majorana mass as shown in Fig.~\ref{fig:mass_gen}. Treating $\Lambda$ as a dimensionful coupling constant instead of a cut-off scale, the neutrino mass becomes
\begin{align}
	m_{\nu} = \frac{m^3_\chi}{2 \pi^2 \Lambda^2} 
						\left(6\ln\frac{m_\chi}{\mu} - 1 \right),
\end{align}
using the dimensional regularization scheme with modified minimal subtraction\footnote{As we shall see later, this is well justified in the UV-complete model which has exactly same loop structure.}, renormalized at the scale $\mu$. We take $\mu$ to be the neutrino mass $m_\nu$, with the incoming momentum $p$ set to zero.

On the other hand, the relic abundance of $\chi$ is determined by the same effective operator\footnote{One has to include the contribution from $\chi^\dag \chi^\dag \nu^\dag \nu^\dag / \Lambda^2$, which involve a different chirality. The interference between the two different chirality contributions is tiny, being proportional to the very small neutrino mass $m_\nu$.}. The DM annihilation cross section reads, up to $v^2_\text{rel}$,
\begin{align}
	\sigma v_\text{rel} = \frac{m^2_\chi}{\pi \Lambda^4}
											  \left(1 + \frac{1}{2} v^2_\text{rel}\right).
\end{align}
We base the computation of the DM relic density on the thermally averaged annihilation cross section $\lan \sigma v_{rel} \ran$, where we include the fact that the number of relativistic degrees of freedom is much smaller for MeV DM as opposed to GeV DM, as described in Ref.~\cite{Griest:1990kh}

\begin{figure}
\centering
\includegraphics[width=0.8\columnwidth]{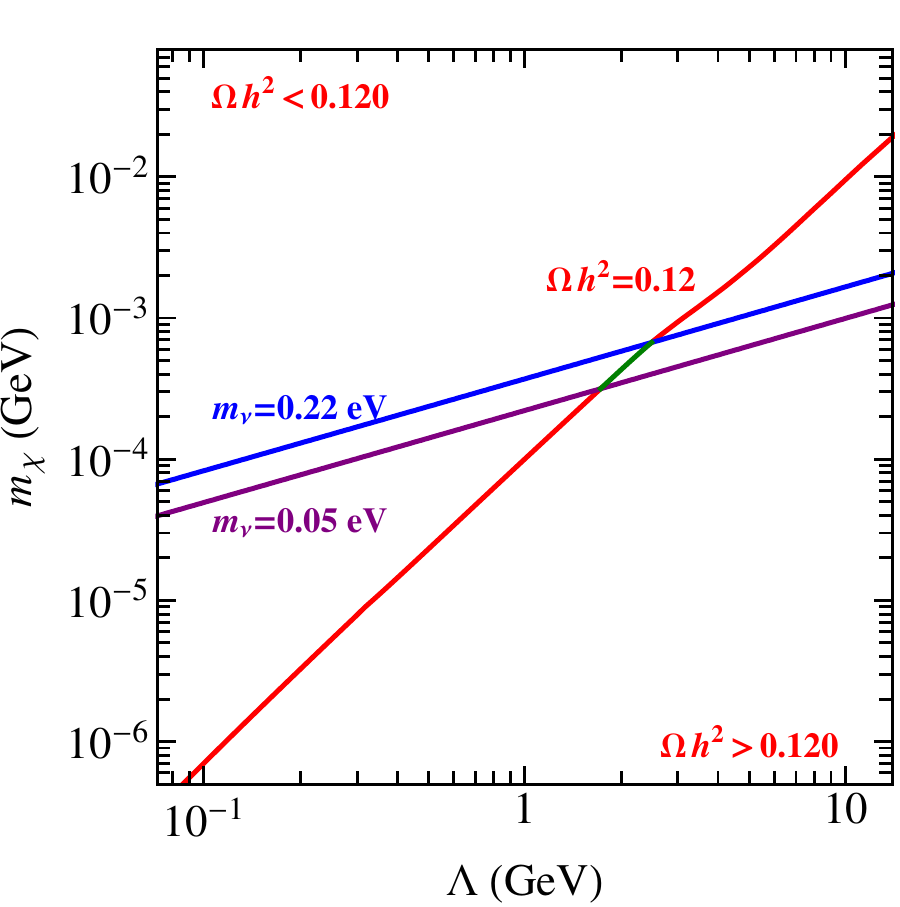}
\caption{Neutrino mass $m_\nu$ and DM relic abundance $\Omega h^2$ as a function of the DM mass $m_\chi$ and the effective DM-neutrino interaction scale $\Lambda$. The red curve corresponds to the correct DM density, while the blue (purple) line refers to the upper~(lower) limit on the heaviest active neutrino mass, as denoted. At the intersection, $m_\chi \approx 0.5$~MeV and $\Lambda \approx 2$~GeV.}
\label{fig:mchi_La}
\end{figure}
Given the observed neutrino mass scale and the DM relic density, the DM mass $m_\chi$ is around the sub-MeV scale while $\Lambda \approx 2$~GeV, as estimated in the previous section. The general relation between the model parameters $m_\chi$, $\Lambda$ and the observables $m_\nu$, $\Omega h^2$ is shown in Fig.~\ref{fig:mchi_La}. The red curve denotes the observed relic abundance $\Omega h^2=0.12$ while the blue (purple) line corresponds to the upper~(lower) limit on the heaviest active neutrino mass. The fact that $\Lambda$ is much smaller than the EW scale justifies the explicit EW symmetry breaking of Eq.~\eqref{eq:effective_L} and it implies the existence of the light particle $\PhiN$ in this scenario.

\section{UV-Complete Toy Model}\label{sec:toy}

\begin{figure}
\centering
\includegraphics[width=0.48\columnwidth]{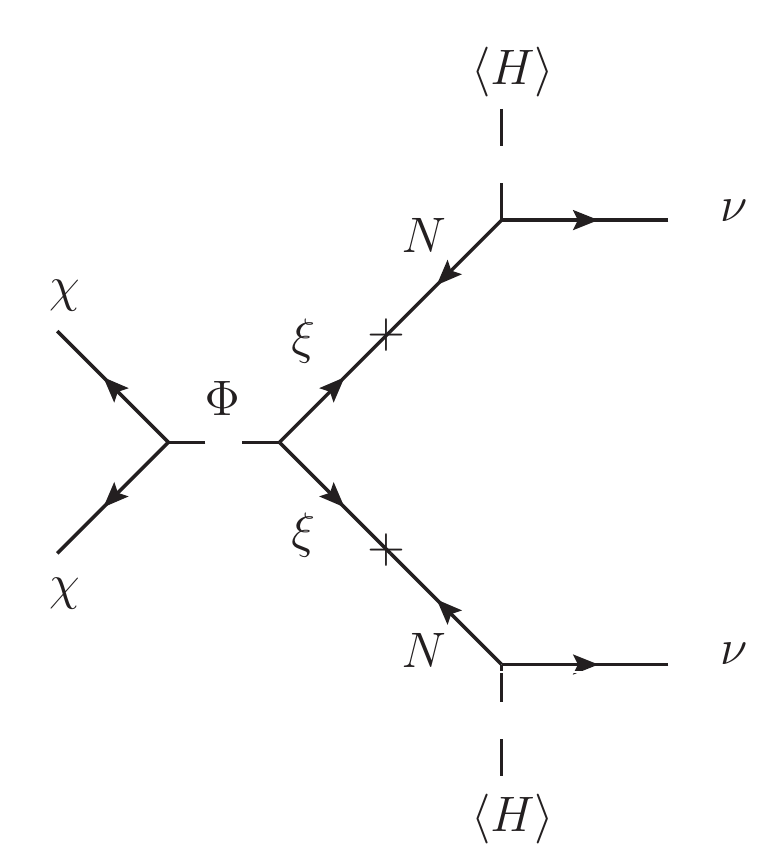}
\caption{Diagram generating the effective DM-neutrino interaction in the UV-complete toy model. For illustration, we only show the diagram with a right-handed $\chi$.}
\label{fig:mass_gen_mod}
\end{figure}

As a final step, we construct a UV-complete toy model that in turn generates the effective Lagrangian and the low energy DM-neutrino interaction, as shown in Fig.~\ref{fig:mass_gen_mod}. The corresponding Lagrangian reads
\begin{align}
\mathcal{L} &\supset 
	 \frac{c_1}{2} \left(\Phichi + \lan \Phichi \ran\right) \chi\chi 
	 + c_2 \PhiN \chi\chi 
	 + c_3 \PhiN^* \xi \xi \nonumber\\
	&+  y LH N   
	 - \mPhichi \Phichi\Phichi^* 
	 - \mPhiN \PhiN\PhiN^* 
	 -\mN N\xi + h.c.,
\label{eq:toy_L}
\end{align}
where $\Phichi$ and $\PhiN$ are scalar fields with lepton number $L = 2$, $N$ and $\xi$ are heavy Dirac neutrinos with opposite $L$. The vacuum expectation value (VEV) $\lan \Phichi \ran$ of $\Phichi$ generates the DM mass $m_\chi = c_1 \lan \Phichi \ran$. In principle, $\Phichi$ could be very heavy compared to $\lan \Phichi \ran$. For instance, $\Phichi$ may couple to another scalar $\phi$ such that $\lan \Phichi \ran = \lan \phi \ran^2/\mPhichi \ll \lan \phi \ran, \mPhichi$, similar to the type-II seesaw mechanism. In other words, the small VEV $\lan \Phichi \ran$ can in this case be triggered by the VEV of $\phi$, suppressed by the heavy $\Phichi$ mass, $\mPhichi$.
Moreover, the massless Majoron from $\phi$ could be removed by gauging $B-L$. The quantum numbers of the various fields are listed in Table~\ref{tab:quantum_number}. The $Z_2$ symmetry is imposed to guarantee the stability of DM and forbid the mixing between DM and the SM neutrino.
\begin{table}[t]
\centering
\begin{tabular}{cccccccc}
\hline
Field         & $L$        & $H$        & $N$   & $\chi$ & $\xi$ & $\Phichi$ & $\PhiN$ \\
\hline
$[SU(2)_L]_Y$ & $2_{-1/2}$ & $2_{1/2}$ & $1_0$ & $1_0$  & $1_0$ & $1_0$     & $1_0$   \\
$L$           & 1          & 0          & -1    & -1     & 1     & 2         & 2       \\
$Z_2$         & +          & +          & +     & --     & +     & +         & +       \\
\hline
\end{tabular}
\caption{Particle content and corresponding quantum numbers in the toy model.}
\label{tab:quantum_number}
\end{table}
Lepton number is spontaneously broken after $\Phichi$ obtains a VEV, giving a Majorana mass to $\chi$. Moreover, $\chi$ induces the mixing between $\Phichi$ and $\PhiN$ such that LNV is transferred to the heavy neutrino $N$ and finally to $\nu$ via the heavy-light neutrino mixing. Integrating out the heavy particles, we obtain the effective DM-neutrino interaction $\chi\chi\nu\nu / \Lambda^2$, with
\begin{align}
	\Lambda = \frac{1}{\sqrt{c_2 c_3}}\frac{\mN \mPhiN}{y v}.
\label{eq:La_fun}
\end{align}
Alternatively, the $\chi$-induced mixing between $\Phichi$ and $\PhiN$ gives rise to a linear term in $\PhiN$ once $\Phichi$ acquires a VEV. This linear term will induce a small VEV of $\PhiN$ as
\begin{align}
	\contraction{\lan \PhiN \ran = c_2}{\chi}{{}{}}{\chi } \lan \PhiN \ran = c_2 \chi\chi /  \mPhiN^2 , 
\end{align}
where $\contraction{}{\chi}{}{\chi} \chi\chi$ represents the $\chi$-loop of mass dimension three. It in turn gives a small Majorana mass term to $\xi$, $ c_3 \lan \PhiN \ran \xi\xi$. The full neutrino mass matrix in the basis ($\nu$, $N$, $\xi$) reads
\begin{align}
  \begin{pmatrix}
    0   & y v & 0   \\
    y v & 0   & \mN \\
    0   & \mN & 2 \frac{c_2 c_3 \contraction{}{\chi}{}{\chi} \chi\chi}{\mPhiN^2} \\
  \end{pmatrix},
\end{align}
which is exactly the inverse seesaw~\cite{Mohapatra:1986bd}. The resulting light neutrino mass will be $m_\nu= 2 c_2 c_3 y^2 v^2 \contraction{}{\chi}{}{\chi} \chi\chi / (\mPhiN^2 m^2_N)$, which implies Eq.~\eqref{eq:La_fun} from Eq.~\eqref{eq:effective_L}. In addition, for $\Lambda \approx 1$~GeV one has $m_N \mPhiN \approx 100\text{ GeV}^2$ if all couplings are of $\mathcal{O}(1)$. This also means that $m_N$ is bounded from above, $m_N \lesssim 100$~TeV since $\mPhiN$ is required to be larger than $m_\chi$, otherwise $\PhiN$ can not be regarded to be heavy to generate the effective operator $\dims / \Lambda^2$. On the other hand, $m_N$ is also bounded from below $m_N \gtrsim 100$~GeV for $y = \mathcal{O}(1)$ due to constraints from EW precision and flavor changing neutral currents data~\cite{delAguila:2008pw, Atre:2009rg, Deppisch:2015qwa}.

The UV-complete toy model satisfies two very important requirements necessary for this mechanism to work: Firstly, LNV arises in the hidden sector, and it is mediated to the SM sector (including right-handed neutrinos) only by the DM particle $\chi$. Secondly, the heavy particles that are being integrated do {\it not} enter the $\chi$-loop, which radiatively induces the light neutrino mass. It is a distinctive feature of the model, setting it apart from the existing literature, for instance Refs.~\cite{Babu:2001ex, Gouvea:2007xp, Angel:2012ug}, where heavy particles exist {\it inside} loops that give rise to radiative neutrino masses. It is this feature that renders our model more predictive.

\section{Conclusions}\label{conclusion}
In this letter, we propose a simple scenario that directly connects the physics of DM and neutrino masses. The introduced operator $\dims / \Lambda^2$ induces a radiative Majorana neutrino mass as well as leads to the DM annihilation. Given the observed DM density and the neutrino mass scale, the DM mass $m_{\chi}$ and the operator scale $\Lambda$ are uniquely fixed to be of order MeV and GeV, respectively. In a UV-complete toy model, we postulate the breaking of lepton number in a hidden sector that is mediated via a DM loop to the visible sector and thus to the light neutrinos, generating the effective Weinberg operator. To our knowledge this has not been explored in the literature but we find this possibility rather suggestive and intriguing; it would for example motivate why lepton number is only slightly broken in the visible sector and the DM loop mediation is quite natural with the presence of a $Z_2$ symmetry to ensure the DM stability, only allowing the DM particle to interact in pairs.

In our letter we focus purely on the relation between the neutrino mass generation and the DM annihilation. As an outlook, we would like to comment on other potential signatures of the model. The DM annihilation cross section is $S$-wave dominated without velocity suppression. This implies a neutrino flux due to ongoing DM annihilation, for example, from the Galactic center. The DM mass and hence the energy scale of the neutrino flux is of order MeV, in the vicinity of the energy threshold of neutrino experiments such as Super-Kamiokande~\cite{Renshaw:2014awa}, KamLAND~\cite{Abe:2011em, Gando:2014wjd}, SNO~\cite{Aharmim:2009gd} and Borexino~\cite{Bellini:2013lnn}. We estimate the expected monochromatic neutrino flux as $\Phi_\nu \approx 300 \, (\text{MeV}/m_\chi)^2$~cm$^{-2}$s$^{-1}$ using the calculation of Ref.~\cite{PalomaresRuiz:2007eu}. Such a flux would give rise to a few events for an exposure of a Mton~$\cdot$~yr.

The fact that the effective scale $\Lambda$ is naturally of order GeV implies the existence of light exotic states. With regard to direct searches at colliders, it is difficult to make a general statement without fully specifying a UV-complete model. For a TeV scale neutrino $N$, $m_\PhiN \approx$~GeV. The scalar $\PhiN$ only couples indirectly through $N$ and is hardly constrained by collider searches. If $\PhiN$ couples to the SM Higgs via $\PhiN^*\PhiN H(H)$, invisible Higgs decays without phase space suppression would be generated. The mass range of the heavy quasi-Dirac neutrino $N$~(and $\xi$) is confined to be 100~GeV $\lesssim m_N \lesssim 100$~TeV with a relatively large heavy-light neutrino mixing. Therefore, the heavy neutrino production cross section could be sizeable at the LHC.

Finally, we would like to point out that the MeV scale DM particle can contribute to the entropy of the universe during the time of Big Bang Nucleosynthesis~(BBN). In our scenario, DM is still in thermal equilibrium with neutrinos after they decouple from the thermal bath around the temperature $T = 2.3$~MeV~\cite{Enqvist:1991gx}. The DM particles subsequently become non-relativistic and transfer entropy to the neutrinos, thereby re-heating the neutrino temperature with respect to that of photons. This leads to a larger number of relativistic degrees of freedom $N_\nu$ during the time of last scattering producing the cosmic microwave background~(CMB). MeV scale Majorana DM coupling to the neutrinos will result in $N_\nu^\text{BBN} = 4$ and $N_\nu^\text{CMB} = 4.4$ \cite{Boehm:2012gr, Nollett:2014lwa}. This is consistent with the BBN observation, $1.8 < N_\nu^\text{BBN} < 4.5$~\cite{Cyburt:2004yc, Agashe:2014kda}, but it is in tension with $N^\text{CMB}_\nu=3.36 \pm 0.34$ from the CMB data alone~\cite{Ade:2013zuv}. Potentially more severe is the constraint from the determination of the primordial Deuterium abundance. The observationally determined value, expressed relative to the Hydrogen abundance, is $D / H = (2.53 \pm 0.04) \times 10^{-5}$~\cite{Agashe:2014kda}. On the other hand, a Majorana DM particle with mass $m_\chi \approx 0.5$~MeV and annihilating to neutrinos as in our scenario would result in a value $D / H \approx 3 \times 10^{-5}$~\cite{Nollett:2014lwa}. Compatibility with observation in this case requires a DM mass $m_\chi \gtrsim 7$~MeV.

The apparent tension of our toy model with astrophysical data can be avoided or at least alleviated in many ways. We would like to highlight two possible options that can be implemented in a more realistic model. Firstly, to realize the observed neutrino mixing pattern and mass-squared differences, at least two generations of $(N, \xi)$ are needed. In this case, one can have a larger DM mass $\gtrsim 7$~MeV while having cancellations in the flavor structure between two contributions to the neutrino masses to keep $m_\nu$ small, such that the CMB and BBN constraints do not apply. The corresponding neutrino mass with $m_\chi \gtrsim 7$~MeV would be around 20~eV, necessitating a $1\%$ tuning by accident or symmetry to achieve the required $m_\nu \approx 0.2$~eV. Secondly, one can add a Dirac component to the DM $\chi$ mass, effectively turning $\chi$ into a quasi-Dirac particle. This would increase the DM mass while the Majorana neutrino mass, proportional to the Majorana DM mass component, could be kept constant. The two solutions, however, will modify the DM and neutrino mass link but they could be implemented in a controllable way via the help of a symmetry such that the connect between DM and neutrino physics still exists.

\subsection*{Acknowledgments}
The authors are especially thankful to John Ellis and Jose Valle for useful comments, and Andr\'e de Gouv\^ea for very detailed and useful comments on the draft. We thank Pedro Schwaller for pointing out the tension with the CMB measurement of the number of relativistic degrees of freedom. W.-C. H. is grateful for the hospitality of the CERN theory group and the AHEP group at IFIC, where part of this work was performed. This work is supported by the London Centre for Terauniverse Studies (LCTS), using funding from the European Research Council via the Advanced Investigator Grant 267352.

\bibliographystyle{h-physrev}
\bibliography{DMnu_mass}
\end{document}